\documentclass[aps,twocolumn,floatfix,bibnotes,altaffilletter]{revtex4}

\usepackage{dcolumn}
\usepackage{color}
\usepackage{amsmath}
\usepackage{here}
\usepackage{graphicx}

\newcommand\leqt[1]{\protect\label{eq:#1}}
\newcommand\reqtn[1]{\ref{eq:#1}}
\newcommand\reqt[1]{(\reqtn{#1})}

\newcommand{\be}{\begin{equation}}
\newcommand{\ee}{\end{equation}}
\newcommand{\bea}{\begin{eqnarray}}
\newcommand{\eea}{\end{eqnarray}}

\newcounter{Fig}

\begin{document}

\begin{sloppy}

\title{Optical Bloch oscillations in periodic structures with metamaterials}

\author{Artur R. Davoyan, Ilya V. Shadrivov, Andrey A. Sukhorukov, and Yuri S. Kivshar}

\affiliation{Nonlinear Physics Center, Research School of Physical
Sciences and Engineering, Australian National University, Canberra, ACT 0200, Australia}


\begin{abstract}
We predict that optical Bloch oscillations can be observed in
layered structures with left-handed metamaterials and zero average
refractive index where the layer thickness varies linearly across
the structure. We demonstrate a new type of the Bloch oscillations
associated with coupled surface waves excited at the interfaces
between the layers with left-handed material and conventional
dielectric.
\end{abstract}


\maketitle

Electron oscillations in the presence of a constant electric field
were predicted by Bloch in 1928~\cite{Bloch:1928-555:ZP}. Such Bloch
oscillations become possible due to beating of the localized
eigenmodes of the structure corresponding to the equidistant
eigenstates of the spectrum known as the Wannier-Stark
ladder~\cite{Wannier:1960-432:PREV}. Experimental verification of
the theory was impossible at that time, since dephasing time of
electrons in crystals is shorter than the period of the electron
Bloch oscillation. Later, electron Bloch oscillations were observed
in semiconductor superlattices~\cite{Waschke:1993-3319:PRL} for
which the period was reduced due to a small mini-band width in the
artificial structure.

Dephasing processes for electromagnetic waves are negligible making
the observation of the optical Bloch oscillations in photonic
systems much easier. The first experimental observation of optical
Bloch oscillations was reported in
Ref.~\cite{deSterke:1998-2365:PRE} for linearly chirped Bragg
gratings. Later, several studies reported the observation of optical
Bloch oscillations in various
structures~\cite{Agarwal:2004-97401:PRL,
Wilkinson:2002-56616:PRE,Sapienza:2003-263902:PRL,Lenz:1999-963:PRL,Pertsch:1999-4752:PRL}.

Recent experimental realization of left-handed
materials~\cite{Smith:2004-788:SCI} has opened up many unique
opportunities to explore novel effects in the structures with
negative refractive index. In this Letter we study, for the first
time to our knowledge, optical Bloch oscillations in one-dimensional
layered structures containing alternating layers of left-handed and
conventional dielectric slabs. We choose the material parameters in
such a way that the average refractive index $\bar{n}$ across pair of the neighboring layers vanishes, thus fulfilling the
condition for the existence of a novel type of the specific
zero-$\bar{n}$
bandgap~\cite{Li:2003-83901:PRL,Shadrivov:2003-3820:APL}. We change
the layer thickness linearly in the structure and observe an optical
analogue of the Wannier-Stark ladder in the eigenmode spectrum, and
the corresponding Bloch oscillations in the resonant transmission
bands. We reveal that in such structures the Bloch oscillations can
be observed in three different regimes. Compared to the photonic
Bloch oscillations in conventional dielectric structures, the
metamaterial structures can support a novel type of the Bloch
oscillations associated with coupling of surface waves at the
interfaces between left-handed and dielectric layers.

We study a one-dimensional layered structure shown schematically in
Fig. 1, where the slabs with the width $b_i$ are made of
metamaterial being separated by a dielectric slab with the width
$a_i$. Variation of the refractive index in the i-th pair of layers can be
described as follows:
\be
n(z)=\left\{
        n_r = \sqrt{\varepsilon_r \mu_r} \; \; \; \; \; \; \;
                z\in(z_i,z_i+a_i) \atop
        n_l = -\sqrt{\varepsilon_l \mu_l} \; \; \; \; \;
        z\in(z_i+a_i,z_i+\Lambda_i)\right.
\ee
where $n_l$ and $n_r$ are the refractive indices of metamaterial and
dielectric, respectively.


\begin{figure}
  \includegraphics[width=\columnwidth]{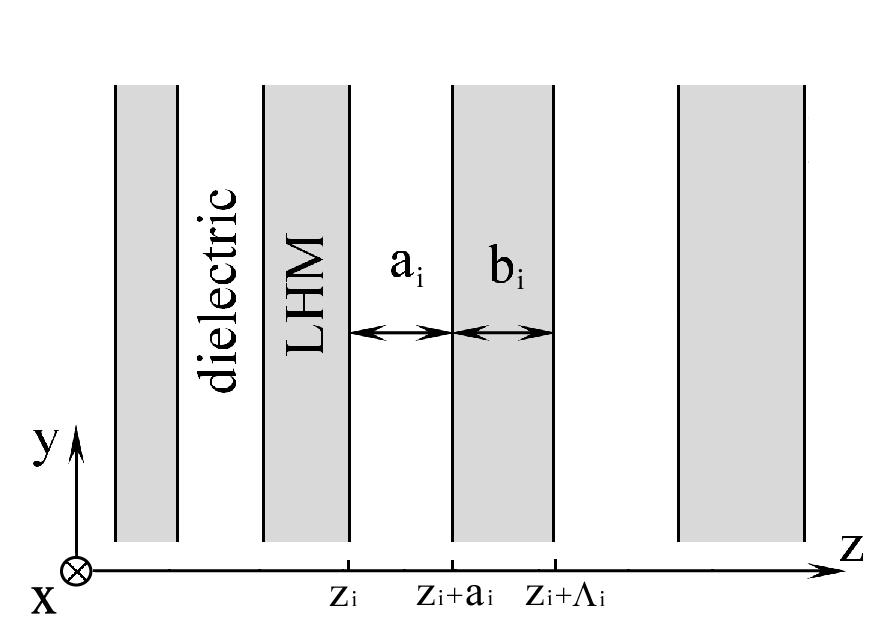}
  \caption{Schematic of linearly chirped one-dimensional photonic crystal
with alternating  layers of left-handed metamaterial and
dielectric.}
\end{figure}

 We consider TE-polarized waves with the electric field
having one component ${\bf E}=(E_x,0,0)$, and waves propagating in
the plane $(y,z)$. In this case, the field distribution can be
described by the Helmholtz equation:

\be\leqt{Helmholtz}
\Delta_2 E_x(y,z)+ n^2(z)E_x(y,z) -\frac{1}{\mu} \frac{d\mu(z)}{dz}\frac{\partial E_x(y,z)}{\partial z}=0,
\ee
where $\Delta_2$ is the two-dimensional Laplacian, and the
coordinates are normalized to $c/\omega$. Firstly we consider periodic structure. Electric field in an infinite one-dimensional {\em periodic structure} can be represented
as a superposition of  Bloch eigenmodes~\cite{Yeh:1988:OpticalWaves},
with the electric field envelopes $U(z+\Lambda)=U(z),$
where $\Lambda$ is the structure period. The dispersion
relation for the Bloch waves is found by the transfer matrix
method~\cite{Yeh:1988:OpticalWaves},
\bea\leqt{dispersion}
2\cos(K_{B}\Lambda) =
    2\cos(k_{zr} a) \cos(k_{zl}b) - \\
    \left(
        \frac{k_{zl}\mu_r}{k_{zr}\mu_l}
        +\frac{k_{zr}\mu_l}{k_{zl}\mu_r}
    \right)
    \sin(k_{zr}a)\sin(k_{zl}b),\nonumber
\eea
where $K_B$ is the Bloch wavenumber, $k_{zl,zr}=\mp\sqrt{n_{l,r}^2-k_y^2}$ and $k_y$ is the normalized propagation constant along the $y$ axis. According to
this relation an infinite stack containing metamaterials exhibits a
non-resonant gap for $\bar{n} \equiv \Lambda^{-1}
\int_{0}^{\Lambda} n(z)dz=0$, where $\bar{n}$ is an average
refractive index of the structure, i.e. $n_{r}a=|n_{l}b|$. This condition is easy to fulfil for negative refractive
index materials.

\begin{figure}
  \includegraphics[width=\columnwidth]{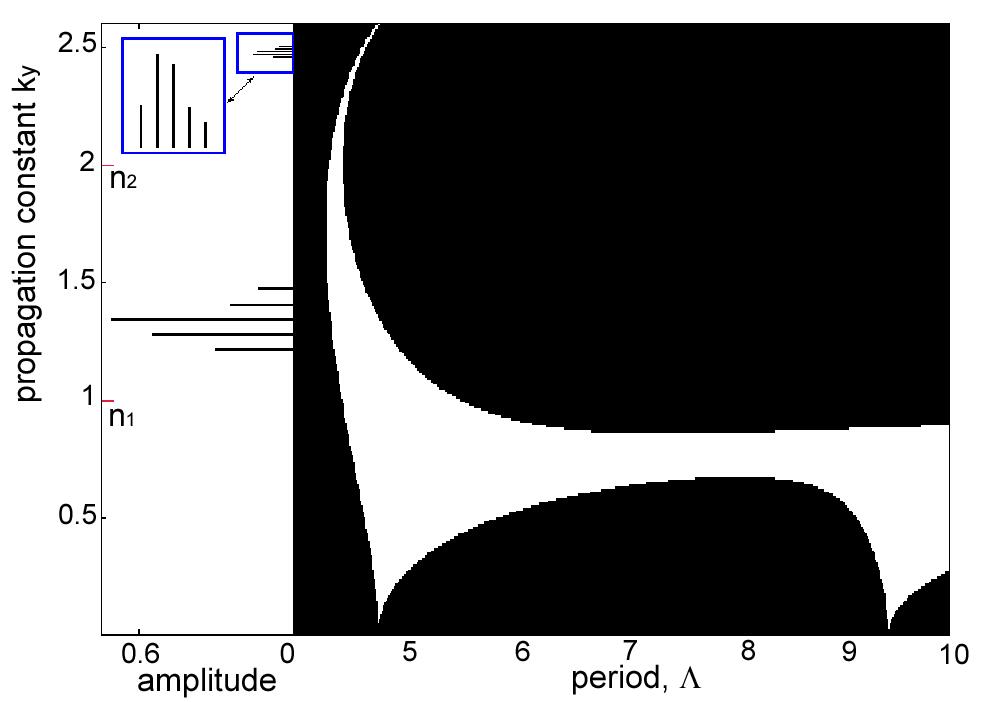}
  \caption{(Color online) Bandgap diagram for the TE-polarized
waves. Black and white areas correspond to gaps and bands,
respectively. Two spectra of the excited Bloch oscillations are
shown on the left. The inset shows a magnified part of the
spectrum.}
\end{figure}


The bandgap diagram of the layered structure is shown in
Fig. 2 for the parameter plane $(\Lambda,k_y)$.
Here we assume that dielectric is vacuum, $\varepsilon_r=\mu_r=1$,
and that it is two times thicker than the second layer, $a/b=2$. We
choose the parameters of the left-handed media as follows:
$\varepsilon_r=-5$ and $\mu_r=-0.8$. This set of parameters allows
surface waves to exist at the interfaces between metamaterial and
vacuum~\cite{Shadrivov:2003-3820:APL}. As follows from
Fig. 2, for the zero $\bar{n}$ structure the
bandgap spectrum differs substantially from the case of conventional
periodic structures made of conventional dielectrics
~\cite{Yeh:1988:OpticalWaves}. Stack with the average zero
refractive index possesses a complete gap with the transmission
resonances~\cite{Li:2003-83901:PRL,Shadrivov:2003-3820:APL} when the
optical path of the wave in either layer of the period coincides
with a half of the wavelength in the corresponding medium. Thus for
the normal incidence ($k_y=0$) transmission is observed only when
$n_r a=n_l b= \pi m$, where $m$ is integer. For the slabs of equal
thickness the regions of the transmission resonances in $(\Lambda,k_y)$
plane degenerate into infinitely thin lines.

We study the propagation of electromagnetic waves in such a layered
structure with zero average refractive index in each pair of layers,
when the thickness of layers is chirped linearly, i.e.
$\Lambda_q=\Lambda_0+ q \delta\Lambda$, where integer $q$ numbers the layers. We are looking for localized
solutions in the structure, and for numerical simulations we
consider a finite stack of layers with perfect metal boundary
conditions, $E(z=0)=E(z=L)=0$, where $L$ is the total length of the
structure. We assume the Gaussian field distribution in the plane
$y=0$ across the layers, and in order to find the electromagnetic
field distribution in the whole stack we look for its eigenmodes by
solving the Helmholtz equation~\reqt{Helmholtz}. Then we decompose
the initial field distribution using the basis of eigenmodes and
find the solution in the whole structure.

To find the eigenmodes of Eq.~\reqt{Helmholtz} we employ the
following discretization scheme~\cite{Sukhorukov:2006-105:IJNM}:
\bea
    \frac{2}{\mu_{m-1}^{-1}+\mu_{m+1}^{-1}}
    \left[
        \frac{U_{m+1}-U_m}{\mu_{m+1}}-
        \frac{U_m-U_{m-1}}{\mu_{m-1}}
    \right]
    \frac{1}{h^2}+\\
    +\frac{\varepsilon_{m-1}+\varepsilon_{m+1}}{\mu_{m-1}^{-1}+
       \mu_{m+1}^{-1}}U_m=k_y^2U_m,\nonumber
\eea
where $z_m=m h$ are the mesh points with the discretization step $h$
and $E_x(y,z)=U(z)e^{-j k_y y}$. Such a discretization scheme
provides an algorithm convergence~\cite{Sukhorukov:2006-105:IJNM},
and it avoids excitation of spurious modes in the structure.

In metamaterials the energy flow $\int [{\bf E}\times {\bf H}]dz$
can be negative, i.e. the energy can propagate in the opposite
direction to the propagation constant ${\bf k_y}$
~\cite{Veselago:1967-2854:SPSS}. Consequently, we determine the
direction of the energy flow of each eigenmode  and choose the sign
of the propagation constant such that the energy flows in the positive
$y$-direction. Decomposition of the initial condition in the plane
$y=0$ in the eigenmode basis is made using the least squares method.

\begin{figure}
  \includegraphics[width=\columnwidth]{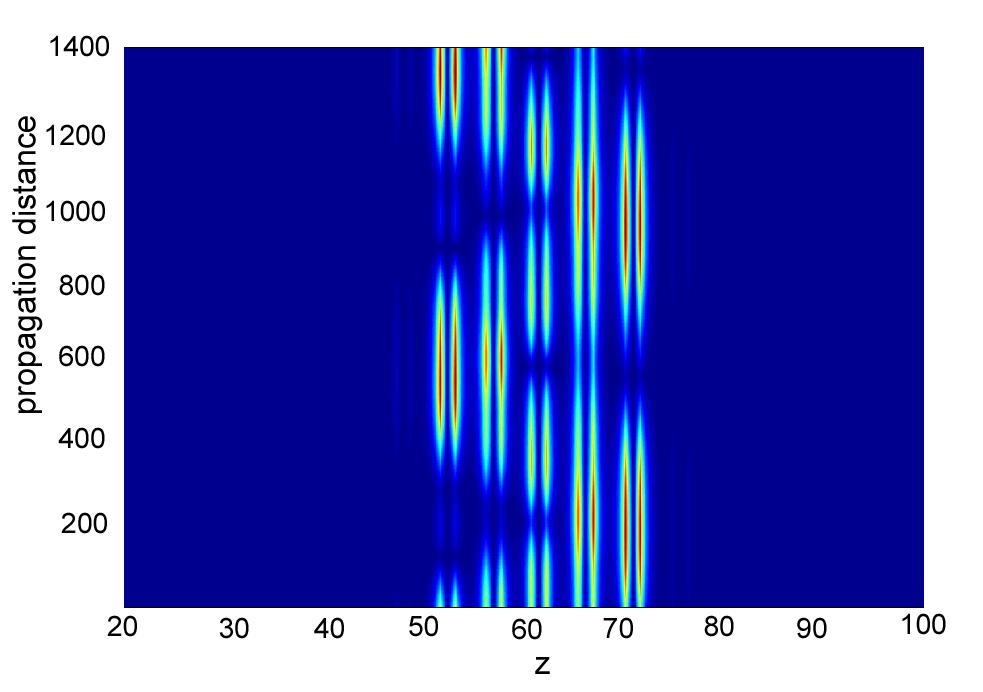}
  \caption{(Color online) Field distribution in the
case of surface-wave-assisted Bloch oscillations. The Wannier-Stark
ladder appears for the propagation constants centered around
$k_{y0}=2.47$, normalized period is $L_y=820$.}
\end{figure}

To find propagation constants (values of $k_y$) which lead
to the Bloch oscillations, we analyze the spectrum of eigenvalues of
this layered structure. The Bloch oscillations are expected to
appear where the spectrum of eigenmodes is equidistant. Practically
for all gradients of a linear ramp we observe several sets of
equidistant states. The equidistant eigenvalues of $k_y$ correspond
to a spatial optical equivalent of the Wannier-Stark ladder which is
associated with the Bloch oscillations.

Spectrum of $k_y$ can be divided into three different regions.
First, when $k_y < n_r < |n_l|$, electromagnetic waves propagate in
both left- and right-handed materials. In the second region,
$n_r<k_y<|n_l|$, waves propagate in metamaterial only being evanescent
in the vacuum layers. In this regime, our structure can be
considered as an array of coupled left-handed waveguides. When
$k_y>|n_l>|n_r$, only surface waves may propagate along the interfaces
separating different materials.

We find that the Bloch oscillations can be observed in all
three regimes of the wave propagation when the corresponding set of
equidistant propagation constants is excited. We consider a stack
containing 36 pairs of metamaterial and dielectric slabs and the
normalized period $\Lambda$ varying from 3.7 to 6. First, we excite
the eigenstates corresponding to the regime of surface waves with
the center of the spectrum at $k_{y0}=2.47$.
Figure 3 presents the intensity distribution for
the electric field which shows clearly spatially periodic
oscillations of the beam position in the structure. The
corresponding spectrum of eigenstates is shown on the left side of
Fig. 2. We note that the beam reconstructs its
shape after each period of oscillations. The field is highly
confined to the interfaces between metamaterial and vacuum,
demonstrating that such Bloch oscillations exist due to interaction
of surface waves in the structure. The distance between
Wannier-Stark eigenstates $\Delta k_y$ defines the period of
oscillations, $L_y=2\pi/\Delta k_y$. For this case, we find
$L_y=820$, and this agrees well with Fig. 3.

\begin{figure}
  \includegraphics[width=\columnwidth]{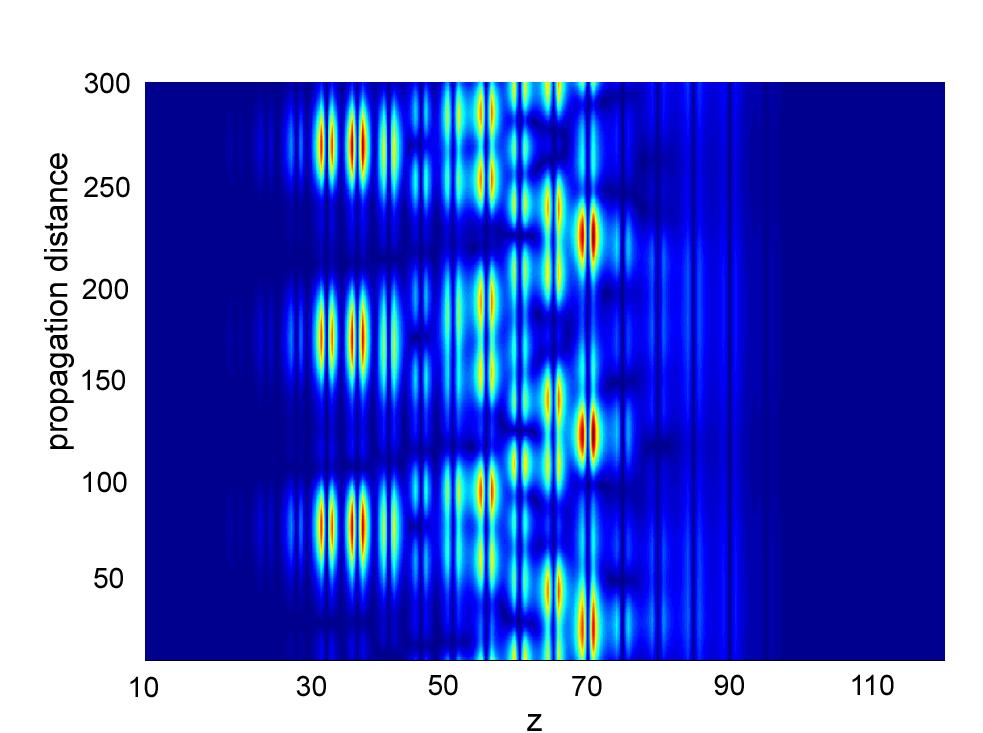}
  \caption{(Color online) Field distribution for the
case of guided waves. The Wannier-Stark ladder appears for the
propagation constants centered around $k_{y0}=1.34$, period is $L_y
\cong100$.}
\end{figure}

Bloch oscillations of the beam with the spectrum corresponding to
the coupled waveguide regime, $n_r<k_y<|n_l|$, are shown in
Fig. 4. The equidistant spectrum of eigenstates
corresponding to the Wannier-Stark ladder is also shown in
Fig. 2 (top, left). We notice that oscillations are strongly
anharmonic, but they are still periodic with the period defined well
by the relation $L_y=2\pi/\Delta k_y$, which is less than the period
of Bloch oscillations associated with surface waves.

The regime of Bloch oscillations corresponding to the waves
propagating in both media can be found in a different structure
with wider transmission resonance. We analyse that the structure
consisting of 36 periods and where the normalized period varies
linearly from 2.5 to 7.5 (corresponding to the period change
gradient $\delta\Lambda= 0.14$). Ratio of the layer thicknesses in
each period is the same as in the previous calculations. We choose
$\varepsilon=-3.6$ and $\mu=-1.11$, preserving the zero average
refractive index of the structure. The calculated field distribution
in this case is shown in Fig. 5, and the center of the
equidistant spectrum appears at $k_{y0} \sim 0.8$.

\begin{figure}
  \includegraphics[width=\columnwidth]{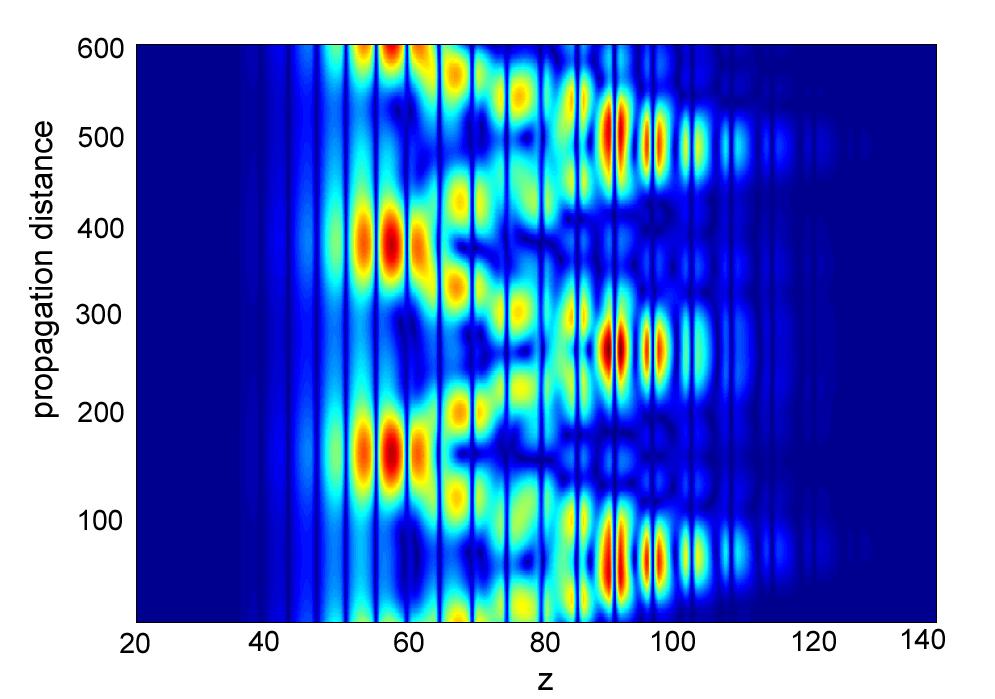}
  \caption{(Color online) Field distribution for the case
when waves propagate in both type of materials. The Wannier-Stark
ladder appears with average propagation constant $k_{y0} = 0.8$,
period of oscillations is $L_y=210$.}
\end{figure}


In conclusion, we have studied the propagation of electromagnetic
waves in layered structures with left-handed metamaterials, and have
demonstrated that linearly chirped structures with zero average
refractive index can support novel types of Bloch oscillations. We
have demonstrated that the  excitation spectra are equidistant,
manifesting a similarity with the optical Wannier-Stark ladder.
Using numerical simulations, we have demonstrated three different
types of the Bloch oscillations, and we have revealed the existence
of unusual oscillations associated with coupled surface waves.

\end{sloppy}
\end{document}